\documentclass{emulateapj}

\shorttitle{MHD simulations of ram pressure stripping}
\shortauthors{Ruszkowski et al.}

\def\gsim{\;\rlap{\lower 2.5pt
 \hbox{$\sim$}}\raise 1.5pt\hbox{$>$}\;}
\def\lsim{\;\rlap{\lower 2.5pt
   \hbox{$\sim$}}\raise 1.5pt\hbox{$<$}\;}

\def\stacksymbols #1#2#3#4{\def\theguybelow{#2}
        \def\verticalposition{\lower#3pt}
        \def\spacingwithinsymbol{\baselineskip0pt\lineskip#4pt}
        \mathrel{\mathpalette\intermediary#1}}
\def\intermediary #1#2{\verticalposition\vbox{\spacingwithinsymbol    
        \everycr={}\tabskip0pt
        \halign{$\mathsurround0pt#1\hfil##\hfil$\crcr#2\crcr
                \theguybelow\crcr}}}

\begin{document}
 
\title{Impact of magnetic fields on ram pressure stripping in disk galaxies}
 
\author{M. Ruszkowski$^{1}$, M. Br{\"u}ggen$^{2}$, D. Lee$^{3}$ and M.-S. Shin$^{4}$}
\affil{$^{1}$Department of Astronomy, University of Michigan, 500 Church Street, Ann Arbor, MI 48109, USA; e-mail: mateuszr@umich.edu (MR)\\
$^{2}$Hamburger Sternwarte, Gojenbergsweg 112, 21029 Hamburg, Germany\\
$^{3}$The Flash Center for Computational Science, The University of Chicago, 5747 S. Ellis, Chicago, IL 60637, USA\\
$^{4}$Department of Physics, University of Oxford, Denys Wilkinson Building, Keble Road, Oxford OX1 3RH, UK\\}

\begin{abstract}
Ram pressure stripping can remove significant amounts of gas from galaxies in clusters and massive groups, and thus has a large impact
on the evolution of cluster galaxies. Recent observations have shown that key properties of ram-pressure stripped tails of galaxies, such
as their width and structure, are in conflict with predictions by simulations. To increase the realism of existing
simulations, we simulated  for the first time a disk galaxy exposed to a uniformly magnetized wind including radiative cooling and self-gravity of the gas. 
We find that magnetic fields have a strong effect on the morphology of the gas in the tail of the galaxy. 
While in the purely hydrodynamical case the tail is very clumpy, the magnetohydrodynamical (MHD) case shows very filamentary structures in the tail. 
The filaments can be strongly supported by magnetic pressure and, wherever this is the case, the magnetic fields vectors tend to be aligned with the filaments.
The ram pressure stripping process may lead to the formation of magnetized density tails that appear as bifurcated in the plane of the sky and resemble the double tails observed in ESO 137-001 and ESO 137-002.
Such tails can be formed under a variety of situations, both for the disks oriented face-on with respect to the ICM wind and for the tilted ones. While this bifurcation is the consequence of the generic tendency for the magnetic fields to produce very filamentary tail morphology, the
tail properties are further shaped by the combination of the magnetic field orientation and the sliding of the field past the disk surface exposed to the wind.
Despite the fact that the effect of the magnetic field on the morphology of the tail
is strong, magnetic draping does not strongly change the rate of gas stripping. For a face-on galaxy, the field tends to reduce the amount of gas stripping compared to the pure hydrodynamical case, and is associated with the formation of a stable magnetic draping layer on the side of the galaxy exposed to the incoming ICM wind. For significantly tilted disks, the situation may be reversed and the stripping rate may be enhanced by the “scraping” of the disk surface by the magnetic fields sliding past the ISM/ICM interface. 
Instabilities, such as gravitational instabilities, undo  the protective effect of this layer and allow the gas to leak out of the galaxy. 

\end{abstract}


\section{Introduction}

Galaxies populate different environments in the universe, ranging from
isolated field regions to dense galaxy clusters. Depending on environment, the
properties of galaxies change: In denser regions the galaxies tend to contain
less neutral gas, show a weaker star formation activity and redder colors
than galaxies in underdense regions. Especially cluster spiral galaxies differ from their counterparts in the field in a number of properties (see \citet{boselli06} for a review): 

\begin{itemize}
\item Cluster spiral galaxies tend to be redder than field spirals. 
\item Cluster galaxies are HI-deficient compared to their field counterparts. The deficiency increases towards the cluster center. Spatially resolved studies reveal that the HI deficiency is caused by a truncation of the gas disks. While the HI disks  of field spirals typically extend beyond the optical disks, the opposite is true for HI-deficient spirals. 
\item  Luminous cluster spirals have, on average, a lower star formation rate. 
The suppression of star formation goes hand in hand with HI deficiency. Spatially resolved studies reveal that the H$\alpha$ disks are also truncated. 
\item An increased 20 cm radio continuum intensity suggests an increase in magnetic field strength by factor of 2-3 compared to field spirals. 
\item Late-type galaxies follow more radial orbits and  tend to have higher velocities than early-type galaxies, which suggests that they are free-falling into clusters. 
\end{itemize}

These observations suggest that one or more processes in cluster environments remove gas from galaxies or make them consume their gas, which leads to a  subsequent decrease of star formation activity and change in color. 
Arguably, the most important process for this effect is stripping of the interstellar medium gas by the ram pressure that the galaxies experience as they move through the intracluster medium (see also the extensive review by \citealt{roediger09} and references therein). This process is called ram-pressure stripping (RPS). In order to interpret the signatures of ram-pressure stripping it is crucial to understand the fate and evolution of the stripped ISM. Other mechanisms that can transform spiral galaxies are `harassment' and tidal interactions. The relative importance of these mechanisms in transforming galaxy morphology,  color, and gas content has been the subject of much debate, with large surveys helping to disentangle the various processes (e.g. \citealt{vandenbosch08}).
Recent deep observations of clusters have revealed very long gas tails in HI  (\citealt{oosterloo05, koopmann08, haynes07}). 
There have also been an increasing number of observations of gas tails in H$\alpha$ (\citealt{kenney08, gavazzi01, yagi07, yoshida08, sun07}) and in X-rays (\citealt{sun06, kim08, sun05,wang04}). Whether RPS can cause these tails and how they survive in the ICM is unknown.  In fact, three of the long gas tails mentioned above have been observed in multiple wavelengths. NGC 4388 has a 120 kpc HI tail (\citealt{oosterloo05}), and an extended emission line region consisting of many faint gas filaments emitting in H$\alpha$ close to the galactic disk (Yoshida et al. 2002, 2004). Another galaxy with multiwavelength emission is ESO 137-001 (Sun et al. 2006, 2007), with a tail detected in both H$\alpha$ and X-ray.
Both ESO 137-001 and ESO 137-002 show double tails \citep{zhang13}. NGC 4438 also may have a very long tail observed in H$\alpha$ (Kenney et al. 2008), while much smaller HI and CO tails have been observed (see Vollmer et al. 2009 and references therein).

The first analytical estimate of RPS appears in a paper by \citet{gunn72} who compare the ram pressure with the galactic gravitational restoring force per
unit area.  Later hydrodynamical simulations of RPS (\citealt{abadi99}, \citealt{quilis00}, \citealt{schulz01},
\citealt{marcolini03,acreman03}, \citealt{roediger06a}, \citealt{roediger06b}, \citealt{roediger07}) confirm that this
analytical estimate is fairly accurate at least for galaxies that move primarily face-on. \citet{kapferer08} and  \citet{kapferer09} have studied the impact of RPS on star formation in spiral galaxies using the SPH code {\tt GADGET}. \citet{tonnesen09} included radiative cooling in the adaptive hydrodynamical grid code {\tt ENZO} in order to investigate the impact of a clumpy, multi-phase interstellar medium. They modeled an inhomogeneous ISM including self-gravity and cooling 
with a resolution of  40 pc and found that low surface density gas is stripped easily from any radius, although the overall mass loss does not differ much from models with a smooth disk. They also find that including cooling results in a very different morphology for the gas in the tail, with a much wider range of temperatures and densities. The tail is significantly narrower in runs with radiative cooling, in agreement with observed wakes. In addition, they expect detectable HI and H$\alpha$ signatures, but no observable X-ray emission for the parameters that they considered.

Evidence for magnetic fields in galaxies and clusters comes mainly from (polarized) radio continuum emission and
Faraday rotation measurements. Spiral galaxies are known to have a regular and a tangled magnetic field
component (\citealt{beck05} and references therein) with typical total field strengths of around $10\mu$G. Magnetic fields in the ICM are inferred from the synchrotron emission of cluster-wide diffuse sources in combination with inverse Compton hard X-ray emission. The magnetic field in clusters can also be determined via the rotation measure (RM) if the field structure and the electron densities are known. Magnetic fields can be studied with background and embedded radio sources as well as with RM-synthesis.  The ICM 
magnetic fields have strengths of the order of 1 $\mu$G and coherence scales of the order of 10~kpc (e.g., \citealt{govoni04, vogt05, bonafede10}). More recently, \citet{kuchar11} measured the power spectrum in the Hydra cluster
using a novel Bayesian method. They concluded that the magnetic field may be $\sim 10$ times stronger and detected no turnover in the magnetic fluctuation power spectrum on scales from 0.3~kpc to 8~kpc indicating that the field correlation length is larger than the outer limit of this range.
However, there is likely a considerable scatter in the values of the magnetic field and the field topologies within individual clusters and between different clusters \citep{carilli}.\\
\indent
One may expect that ambient magnetic fields get draped around the galaxies as they move through the ICM when the dominant correlation length of the magnetic field is comparable to or larger than the size of the galaxy. This effect of magnetic draping has been studied by \citet{lyutikov06} and \citet{ruszkowski07,ruszkowski08} for the cases of radio lobes in galaxy clusters. It has also been pointed out that spiral galaxies that move through the ICM sweep up ambient magnetic field to form a dynamically important sheath around them. As pointed out, e.g., by \citet{pfrommer10} and \citet{dursi08}, the magnetic field strength in the boundary layer is set by a competition between sweeping up and slipping of field lines, resulting in a magnetic energy density ahead of the galaxy that is comparable to the ram pressure experienced by the galaxy (Interestingly, the draping effect occurs even when the correlation length of the magnetic field is somewhat smaller than the size of the obstacle; Pfrommer, priv. comm.). For typical conditions in the ICM of $n_{\rm ICM} \sim 10^{-4}$ cm$^{-3}$ and galaxy velocities $v_{\rm gal}\sim 1000$ km s$^{-1}$, this leads to a maximum field strength in the bow of the galaxy of $B =\sqrt{8\pi \rho v_{\rm gal}^2}\sim 6.5\, \mu$G.
Draping would modify the process of stripping, and the magnetic fields may prevent Kelvin-Helmholtz instabilities. The effects of magnetic fields on the ram-pressure stripping process have hardly been studied. A related work by \cite{otmianowska03} used ZEUS3D in combination with results of sticky particle simulations to study mainly the polarization properties of disk galaxies, and considered only disk fields. Their simulation, however, neglected the dynamical coupling of the magnetic fields to the gas.

\section{Methods}
\subsection{Numerical techniques}

The simulations were performed with a proprietary version of the {\tt FLASH} v4 code. {\tt FLASH} code is a
multi-physics parallel adaptive mesh refinement MHD code. Magnetic fields were computed using the unsplit staggered mesh (USM)
method \citep{lee09,lee13}. The USM module is based on a finite-volume, high-order Godunov scheme (3rd order PPM) with 
constrained transport (CT). Major strength of the USM algorithm is that it preserves vanishing divergence of the magnetic field
to very high accuracy. \\
\indent
We use a hybrid self-gravity solver. It solves for the potential on the highest uniformly refined level using a parallel FFT method.
The potential in regions refined beyond that level of refinement is computed using an iterative V-cycles method. This hybrid approach noticeably improves code performance. In order to prevent artificial fragmentation we modified the equation of state by introducing a pressure floor at $P\approx 3G\Delta x^2\rho^2$, where $\Delta x$ is the resolution, $G$ is Newton's constant, and $\rho$ is the gas density \citep{truelove,agertz}. 
We point out that, just as in \citet{truelove}, the simulations presented here were evolved for a few (galactic) dynamical times.
This is enough to resolve the tendencies of the gas to fragmentation (that enhances stripping rate; \citet{tonnesen09}), but it may not be sufficient to fully capture
turbulence driven by gas self-gravity on scales smaller than a Jeans length or a few resolution elements, or to resolve the internal structure of the self-gravitating clouds \citep{federrath11}. However, the kinetic energy of the cold clumps formed in the disk is dominated by their orbital motion in the dark matter potential of the galaxy rather than by turbulent kinetic energy. We also note that the ICM magnetic field that enters the disk is already dynamically important and significantly amplified by processes other than self-gravity-driven turbulent dynamo.\\
\indent
Radiative cooling was computed using Sutherland \& Dopita cooling functions \citep{suth93} and extended to temperatures below $10^4$K using cooling function of \citet{dalg72} truncated below 300 K similarly to, e.g., \citet{tonnesen09}. We assumed solar metallicity, introduced
a density floor at $10^{-29}$ g cm$^{-3}$, and used an ideal gas equation of state with an adiabatic index of $\gamma=5/3$. The mean molecular weight was allowed to vary as a result of the changes in ionization level and was computed self-consistently from the Sutherland \& Dopita tables (for simplicity, for $T<10^4$ K, we assumed lowest temperature value from these tables). Whenever the cooling timescale
becomes shorter than the Courant timestep, we applied the subcycling method (e.g., \citet{anninos,proga,walch}) to the cooling sink term in order to accelerate the computations.

\subsection{Initial conditions}

\begin{table*} \label{params}
\begin{center}
\caption{Potential parameters}
\begin{tabular}{@{}cccccccccc}
\hline 
$\rho_{\rm dm0}$          & $\rho_{d}$          & $M_{b}$                 & $r_{{\rm dm}}$ & $r_{b}$ & $r_{t}$ & $r_{d}$ & $z_{t}$ & $z_{d}$ \\ 
$[10^{-24}$ g cm$^{-3}$]   & [10$^{-24}$ g cm$^{-3}$]  & [$10^{10}$M$_{\odot}$]  &  [kpc]   & [kpc] &  [kpc] & [kpc] & [kpc] &  [kpc]    \\ 
\hline
\hline
   $6.5\times 10^{-1}$  & $2.46\times 10^{1}$  & $3.5$ & $9$   &  $0.8$   & $4.26\times 10^{1}$  & $4.83$  & $4.41$  &  $5\times 10^{-1}$    \\
\hline
\end{tabular}
\end{center}
\end{table*} 

In order to set up the initial distribution of gas density and temperature, we follow a method similar to that described in \citet{mel08}.
The distribution of gas density is made up of disk and halo components. The halo was assumed to be isothermal
with a temperature $T_{h}=8\times 10^{6}$K and a density given by

\begin{eqnarray}
\rho_{h}(r',z') = \rho_{h0}\exp\left[-(\phi(r',z')-\phi(0,0))/c_{s}^{2}\right],\\  \nonumber
\end{eqnarray}

\noindent
where $c_{s}$ is the isothermal sound speed, $\phi(r',z')$ is the total gravitational potential, $\rho_{h0}=2.17\times 10^{-27}$g cm$^{-3}$, and $r'$ and $z'$ are the radial and vertical coordinates measured along the disk plane and the rotation axis of the galaxy, respectively. The distinction between $(r',z')$ and $(r,z)$, where $r$ is measured in the horizontal plane of the simulation box and $z$ along the vertical direction (i.e., along the tall side of the box), is important for the cases where the disk is tilted with respect to the direction of the ICM wind ($+z$ direction). This gas density distribution guarantees that the halo is in hydrostatic equilibrium. \\
\indent
The gravitational potential of the galaxy consists of three components: dark matter halo, stellar bulge, and flattened stellar disk.
The parameters of these components are summarized in Table 1. These components are either analytical or are precomputed
and read in at the code initialization stage.
While the gravitational potential contribution from the gas was neglected at the initialization stage, the gas self-gravity was included at runtime. In the absence of 
radiative cooling the gas configuration quickly settled to an equilibrium state. The dark matter density distribution was assumed to follow a non-singular isothermal sphere

\begin{eqnarray}
\rho_{\rm dm}(r')=\frac{\rho_{\rm dm0}}{1+\left(\frac{r'}{r_{\rm dm}}\right)^{2}}.\\  \nonumber
\end{eqnarray}

\noindent
The stellar bulge potential contribution is described by the Hernquist profile (Hernquist 1990)

\begin{eqnarray}
\phi_{\rm b}(r')=-\frac{GM_{b}}{r'+r_{b}}.\\  \nonumber
\end{eqnarray}

\noindent
We computed the gravitational potential contribution corresponding to the flattened stellar disk by solving the 
Poisson equation. The density distribution in this component is given by the flattened King profile

\begin{eqnarray}
\rho_{\rm d}(r',z') = \frac{\rho_{d}}{\left[1+\left(\frac{r'}{r_{d}}\right)^{2}+\left(\frac{z'}{z_{d}}\right)^{2}\right]^{3/2}}.\\ \nonumber
\end{eqnarray}

\noindent
Stellar mass distribution was truncated for $(r,z)$ such that $\sqrt{(r'/r_{d})^{2}+(z'/z_{d})^{2}}>r_{t}/r_{d}=z_{t}/z_{d}$.\\
\indent
The gas density and pressure of the isothermal halo were computed in the entire computational domain. The gas density in the disk 
was given by a sum of four components each of which was assumed to follow the functional form suggested by \citet{wol03}

\begin{eqnarray}
\rho(r',z')_{i} = \frac{\Sigma_{d}}{2z_{i}}\exp\left( -\frac{r_{0i}}{r'} - \frac{r'}{r_{i}}-\frac{|z'|}{z_{i}} \right)\\ \nonumber
\end{eqnarray}

\noindent
The values of $r_{0i}$, $r_{i}$, and $z_{i}$ for all gas components ({\tt H}$_{\rm in}$, {\tt H}$_{\rm out}$, {\tt H}$_{2}$, and {\tt HII})  
are given in Table 2. The names of these gas components are not to be taken literally but merely serve as a way of setting up the initial gas density distribution in the disk. The transition between the inner {\tt H}$_{\rm in}$ and outer {\tt H}$_{\rm out}$ neutral hydrogen density distributions was smoothed out
according to $\rho = g(x)\rho_{\rm in}+[1-g(x)]\rho_{\rm out}$, where 

\[
  g(x)=\left\{
  \begin{array}{l l}
     1                             & \quad \text{if $x\le 0$}\\
     1-1.5 x^{2}+0.75x^{3}         & \quad \text{if $0<x\le 1$}\\
     0.25(2-x)^{3}                 & \quad \text{if $1<x<2$}\\
     0                             & \quad \text{if $x\ge2$}\\
  \end{array} \right.
\]

\noindent
where $x=(r'-13{\rm kpc})/{\rm 1 kpc}$.
\begin{table}[h!] \label{params}
\begin{center}
\caption{Density distribution}
\begin{tabular}{@{}lccccccc}
\hline 
Component                 &  $\Sigma_{d} [M_{\odot}/{\rm pc}^{2}]$  &  $r_{0i} [{\rm kpc}]$    &  $r_{i} [{\rm kpc}]$ &  $z_{i} [{\rm pc}]$   \\ 
\hline
\hline
 {\tt H}$_{\rm in}$      &     $2.38\times 10^{1}$          &    $1.0$                  &  $1.0\times10^{3}$            & $5\times 10^{2}$         \\
 {\tt H}$_{\rm out}$     &     $1.71\times 10^{3}$          &    $1.0\times 10^{1}$     &  $4.0$                      & $5\times 10^{2}$         \\
 {\tt H}$_{2}$           &     $1.72\times 10^{2}$          &    $3.3$                  &  $2.89$                     & $5\times 10^{1}$         \\
 {\tt HII}               &     $4.17$                       &    $0.0$                  &  $3.0\times 10^{1}$         & $5\times 10^{2}$         \\
\hline
\end{tabular}
\end{center}
\end{table} 
The density distribution given by Equation 5 was truncated using a tapering function 

\[
  h(r) = \left\{
  \begin{array}{l l}
    1                                       & \quad \text{if $r'\le r_{1}$}\\
    \left[1+\cos(\pi (r'-r_{1})/(r_{2}-r_{1}))\right]/2 & \quad \text{if $r_{1}<r'<r_{2}$}\\
    0                                       & \quad \text{if $r'\ge r_{2}$}\\
  \end{array} \right.
\]

\noindent
where $r_{1} = 20$ kpc and $r_{2} = 25$ kpc.\\
\indent
The boundary conditions for the density and temperature of the disk component were set to match those of the halo gas at 
$z_{m}=40$ kpc above the disk midplane. In order to compute the gas temperature in the disk, we integrated the equation of hydrostatic equilibrium 
along the $z'$-direction from $z'=z_{m}$ to $z'=0$ kpc (disk midplane). This approach did not require us to specify the centrifugal force acting on the gas. 
In order to compute the final distribution of densities and temperatures in the entire computational domain, we either selected the precomputed
halo gas densities and temperatures or chose the disk gas properties 
depending on which component dominated the gas pressure. The regions where the disk gas pressure dominated
correspond to the passive scalar values equal to unity, while in the regions dominated by the halo pressure the passive scalar was set to zero. 
We note that the presence of hot gas halo is important only in the initial stages of the simulation, i.e., before the incoming ICM 
wind has the chance to sweep the gas.
Once the gas density and temperature distributions have been computed, we calculated the gas velocity required to keep the gas in 
rotational hydrostatic equilibrium from 

\begin{eqnarray}
v_{\phi}^{2}=v_{c}^{2}+\frac{r'}{\rho}\frac{\partial P}{\partial r},\\ \nonumber
\end{eqnarray}

\noindent
where $v_{c}^{2}=r'\partial\phi/\partial r$. While the disk component of the gas follows this velocity curve, the halo component was assumed to be
initially static. Models corresponding to the values listed in Table 1 and 2 have disk gas mass of 
$\sim 3\times 10^{10}$ M$_{\odot}$, stellar mass in bulge and disk $\sim 1.3\times 10^{11}$ M$_{\odot}$, and halo mass $\sim 1.3\times 10^{12}$ M$_{\odot}$ assuming halo truncation radius of 150 kpc.\\
\indent
The computational domain was 120 kpc$\times$120 kpc$\times$240 kpc. The center of the disk of the galaxy was
was located at $z=52.5$ kpc. We used static mesh refinement. The domain was fully refined 
for $z\le 81$ kpc and had one or two levels of refinement less throughout the rest of the volume. This ensured that
the disk was always maximally refined. It also allowed us to take advantage of the fast gravity solver in the most of the volume providing 
noticeable improvement in the code performance. The maximum refinement level was 6 (low resolution) and 7 (high resolution).
These refinement levels correspond to maximum spatial resolutions of 234 pc and 117 pc, respectively. We compared the disk gas mass (defined as having passive scalar equal to unity) in the initial state for our two resolutions and found that the masses differed by only 1.4\%.
While the refinement blocks have the same aspect ratio as the whole computational domain, each zone is cube-shaped. Each block contained $16\times16\times16$. We note that in the adaptive-mesh refinement tests the entire domain got refined
in just one shock crossing time. In that mode, the use of the fast hybrid gravity solver was not warranted as the gravity solver was accelerated only on 
a small number of blocks that had a low refinement level. Therefore, we decided to use highly refined static mesh in conjunction with the hybrid gravity solver.\\
\indent
The equilibrium gas density, temperature, and velocity were precomputed on a grid with 25 pc resolution in the z'-direction and 50 pc in the radial direction.
In order to set up the initial state, we used a gravitational potential precomputed on a logarithmic grid over a range from 15.1 pc to 515.8 kpc in both $r'$ and $z'$ using 400 logarithmically spaced points in each direction.
These initial conditions were mapped onto the simulation grid at the code initialization stage. 
The tabulated values of the gravitational acceleration due to the static potential were also computed in advance at very high resolution. These tables
were then used to interpolate the gravitational acceleration at runtime. In addition to the acceleration due to the static potential, the gas was
also allowed to self-gravitate. We verified that the initial conditions were globally stable, and the system quickly settled to an equilibrium state in the absence of radiative cooling. When cooling was present the disk became Toomre-unstable as expected.

\subsection{Boundary conditions}
We used isolated boundary conditions for the hybrid multigrid+PFFT gravity solver (using Dirichlet boundary for the multigrid).
That is, we set {\tt grav}$\_${\tt boundary}$\_${\tt type} to ``isolated'' and {\tt mg}$\_${\tt boundary}$\_${\tt type} for all boundaries to ``dirichlet'' in the {\tt FLASH} parameter file. Outflow boundary conditions for the gas are used 
on all but one boundary. On the inflow boundary we assumed that the gas had the temperature of $T=7\times 10^7$ K, a density
of $\rho_{\rm wind}=0.95\times 10^{-27}$ g cm$^{-3}$, and a maximum inflow speed of $v_{\rm max}=1.3\times 10^{8}$cm s$^{-1}$. 
The inflow speed $v_{\rm in}(t)=f_{\rm in}(t)v_{\rm max}$ was increased gradually from 0 to the maximum speed according to the following time profile\\

\[
  f_{\rm in}(t)=1- \left\{
  \begin{array}{l l}
     1-1.5 x^{2}+0.75x^{3}         & \quad \text{if $x\le 1$}\\
     0.25(2-x)^{3}                 & \quad \text{if $1<x<2$}\\
     0                             & \quad \text{if $x\ge2$}\\
  \end{array} \right.
\]

\noindent
where $x=t/\Delta t$ and $\Delta t\approx 59$ Myr. The direction of the inflow was along the $z$-axis and parallel to 
the long side of the computational box. The magnetic field injected at the boundary was in the $x-y$ plane such that $|B_{x}|=|B_{y}|\approx 2\mu$G. 
This yields a Mach number of the inflow of $M=v/c_s \approx 1$, a plasma $\beta=p/(B^2/8\pi)\approx 21\gg 1$ and 
an Alfvenic Mach number of $M_{A}=M\sqrt{\beta\gamma/2}\approx 4$. Note that the ratio of the wind ram pressure to the magnetic pressure 
$P_{\rm ram}/P_{B}=2M_{A}^{2}\approx 37\gg 1$. 
The ICM gas density and temperature adopted in our model are representative of those in the observed clusters. For example, our ICM wind density of $\sim 0.5\times 10^{-3}$ cm$^{-3}$ and temperature $\sim 7$ keV are close to those detected in Coma cluster at $r\sim 0.6$ Mpc from the center \citep{simon13}. At this radius, \citet{bonafede10} find a magnetic field of $\sim 2\mu$G with significant scatter. However, somewhat larger values of the magnetic field are also consistent with observations given the degeneracy of the observationally inferred magnetic field and the slope of the radial distribution of the magnetic field discussed in that work. Our adopted value is also consistent with cluster-averaged lower limit of $\sim 1.7\; \mu$G based on the Fermi data \citep{tesla} for that object.
Since in general there are no available strong constraints on the level of the magnetic field in the outer cluster parts, one can also attempt to extrapolate the existing radial profiles to the outer cluster parts. For example, the above values of the density and temperature correspond to $\sim0.7$Mpc in the well-studied Perseus cluster \citep{simon11}. For the lower limits on the central magnetic field in this object of 4 to 9 $\mu$G \citep{aleksic}, and assuming B-field profile shapes adopted in that work, one can obtain lower limits on the B-field of 1 to 2$\mu$G at $\sim0.7$ Mpc, which is consistent with the values adopted in the simulations.
Further argument against adopting smaller values of the magnetic field in the simulations is purely numerical. For weaker fields, the width of the draping layer becomes thinner and more difficult to resolve.\\ 
\indent
The choice of field geometry was motivated primarily by simplicity. 
The actual topology of the magnetic field in real galaxies is either not known, or its description would require us to introduce additional free 
parameters to the model. 
The morphology of the internal steady-state field is likely to be governed by a galactic dynamo, the study of which is beyond the scope of our current work.
The simulated steady-state internal magnetic fields may also depend on other factors (e.g., the presence of winds) that affect the relative strength of the 
global toroidal and poloidal galactic field components in a differentially rotating galaxy. 
Thus, we decided to start with the simplest setup in which the effect of the external fields can be simulated. A similar approach was adopted 
by \citet{pfrommer10} who considered only uniform external fields and approximated galaxies as nonrotating high density clouds. We extend their work by including galactic 
rotation, hot galactic halo, radiative cooling, and self-gravity. We defer the investigation of the impact of internal galactic fields on the ram 
pressure stripping process to a future publication.\\
\indent
The simulations were performed on the {\it Pleiades} machine at NASA Ames on 1024 processors. 
Main MHD runs required $\sim 200,000$ CPU hours to advance to 0.7 Gyr per run, while their hydro counterparts took roughly $\sim 50,000$ CPU hours to reach 0.7 Gyr. Highest resolution runs used to infer the evolution of stripping rate required $\sim 150,000$ CPU and $\sim 60,000$ CPU in the MHD and hydro cases, respectively. In these runs, evolved to $\sim 0.45$ Gyr, the galactic disk was refined to the highest maximum resolution and the mesh was progressively derefined away from the disk, thus reducing the computational cost per unit of simulation time. 

\section{Results}

\begin{figure} 
\begin{center}
\includegraphics[scale=0.25]{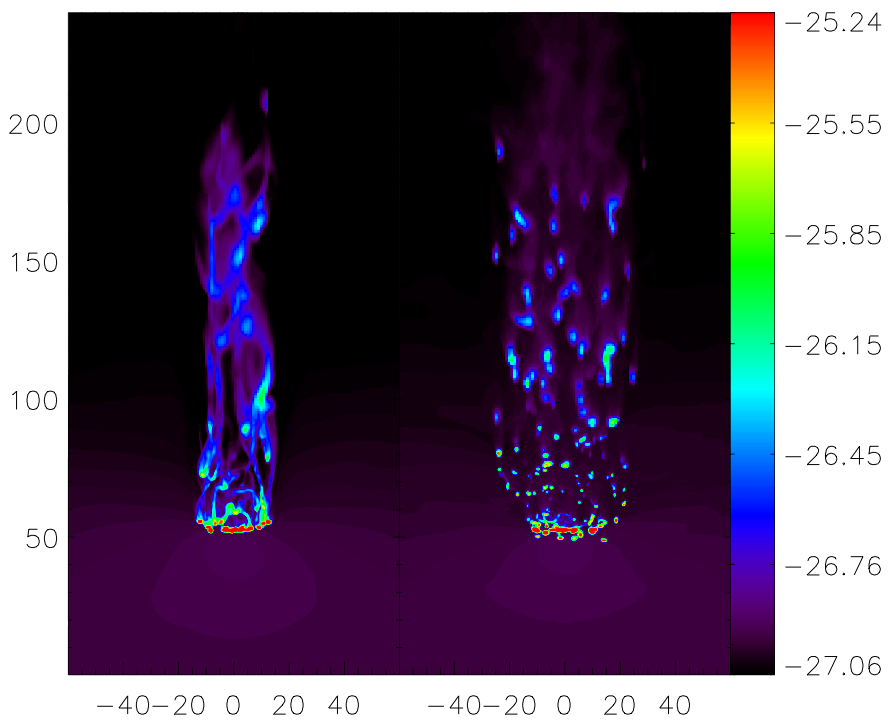}  
\end{center}
\caption{Gas density projected along the line of sight (in log space). Left:
MHD case. Right: non-magnetic case. Both cases are for $t=0.7$ Gyr. The difference between tail morphologies in these 
two cases is striking.\\}
\label{SurfaceDensity}
\end{figure} 

\subsection{Density tail morphology and gas stripping rates in MHD and hydro runs}
 
Figure~\ref{SurfaceDensity} shows the gas 
density projected along the line of sight ($y$-direction) 0.7 Gyr after the start of the wind. The left panel corresponds to the 
MHD case and the right panel to the non-magnetic case. Both panels correspond to the end stages in the simulation and show the case of the galaxy oriented face-on with respect to the ICM wind. The difference between tail morphologies in these 
two cases is striking. While the purely hydrodynamical case
the tail is very clumpy, the MHD case shows very filamentary structures. As a result of this large-scale inhomogeneity in the MHD case, 
the surface density shows more variation with the orientation of projection in the sense that the number of filaments seen in projection depends on the line of sight.

\begin{figure*} 
\begin{center}
\includegraphics[scale=0.4]{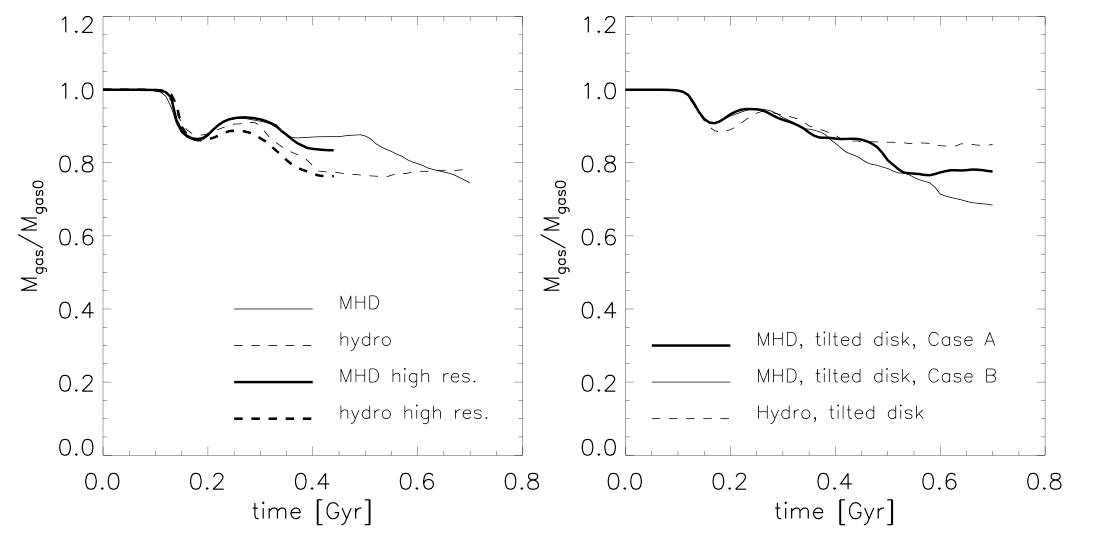}  
\end{center}
\caption{Left: time dependence of the galactic disk gas mass normalized to its initial value for face-on disk cases.
Solid and dashed lines are for the MHD and hydro cases, respectively.
Thick lines correspond to the higher resolution cases. Right: time dependence of the galactic disk gas mass normalized to its initial value for tilted disk cases. 
Solid and dashed lines correspond to the MHD and hydro cases, respectively. In MHD Case A the magnetic field in the incoming ICM wind is $B=[B_{x},B_{y},0]$ and in Case B  the field is $B=[B_{x},-B_{y},0]$, while the wind is always along the z-direction. \\}
\label{DiskGasMass}
\end{figure*} 
\indent
Figure~\ref{DiskGasMass} quantifies the amount of ram pressure stripping of the galactic gas. Shown as a function of time is the
disk gas mass normalized to its initial value. The mass in the disk is computed 
within the region of width and depth equal to $\Delta x=50$ kpc, and height $\Delta z=+/-$6 kpc centered on the disk mid-plane,
where the galactic passive scalar values are greater than 0.25. The disk gas mass is then obtained by integrating 
the gas density multiplied by the passive scalar values in this volume. 
Solid and dashed lines are for the MHD and hydro cases, respectively.
Thick lines in the left panel correspond to the higher resolution cases. In the right panel, Case A corresponds to the magnetic field in the incoming ICM wind  $B=[B_{x},B_{y},0]$ and in Case B  the field is $B=[B_{x},-B_{y},0]$, while the wind velocity is always along the z-direction.
\begin{figure*} 
\begin{center}
\includegraphics[scale=0.4]{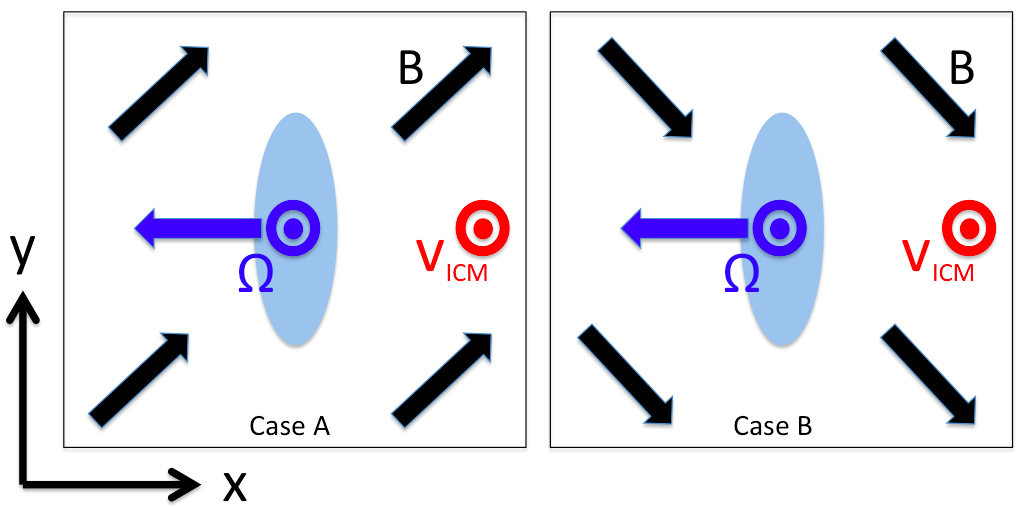}  
\end{center}
\caption{Schematic representation of the initial conditions as seen along the $-z$ direction along the tall side of the computational domain (note that, in this projection, the galactic spin vector $\Omega$ shown in blue has a $-x$ component and a $+z$ component). The magnetic field is shown with black arrows, and the ICM velocity vector points in the $+z$ direction. Case A ($B=[B_{x},B_{y},0]$) and B ($B=[B_{x},-B_{y},0]$) are shown in the left and right panels, respectively.\\}
\label{f3}
\end{figure*} 
Figure~\ref{f3} shows a schematic sketch of the initial conditions as seen along the $-z$ direction. Note that, in this projection, the galactic spin vector has a $-x$ component and a $+z$ component.
In the hydrodynamical cases, 
the prescription for calculating the retained mass leads to results that are entirely independent of the passive fluid threshold as long as 
it is less than $\sim 0.5$. The MHD case is even less sensitive to this threshold.
We point out that threshold values close to unity may underestimate the
volume occupied by the galactic gas, which in turn would lead to the overestimate of the amount of gas stripping.
We also note that Tonnesen \& Bryan (2009) use the same threshold for the passive scalar to identify 
the ram pressure stripping tails in their simulations, and also conclude that the results remain unchanged 
if a threshold of 0.1 is used instead. The rebound in $M_{\rm gas}$ around 0.2 Gyr is due to the gas fallback after the initial episode of ram pressure stripping. 
This effect occurs because some gas temporarily leaves the region inside which the measurement of the galactic gas mass is taken (see also \citealt{roediger06a}).\\
\begin{figure} 
\begin{center}
\includegraphics[scale=0.25]{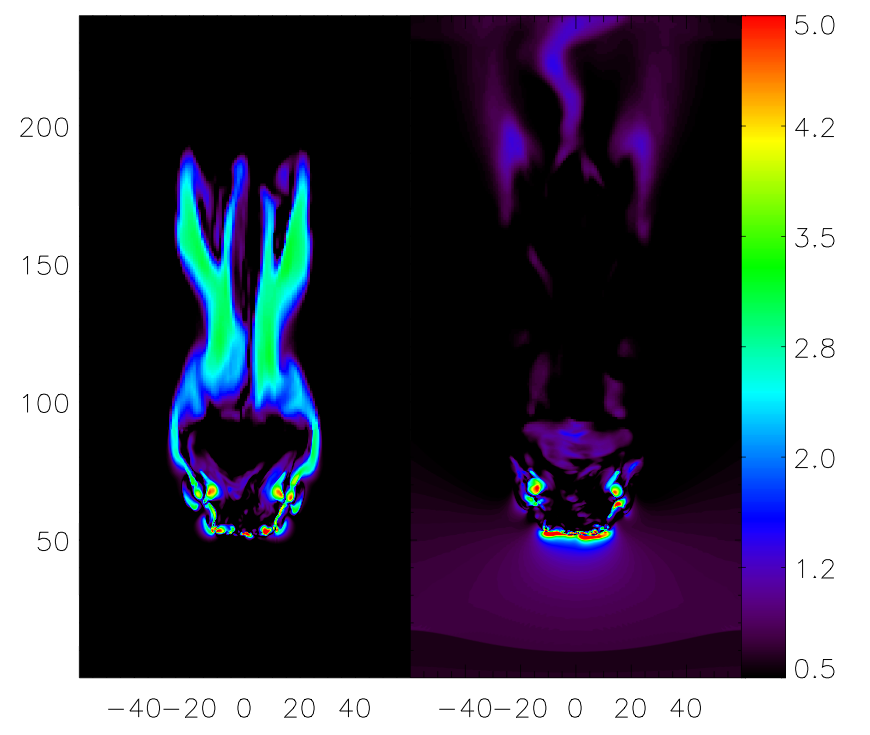}  
\end{center}
\caption{Cross sections ($y=0$) through the computational domain showing the absolute value of the vertical magnetic field $B_{z}/\sqrt{4\pi}$ (left) and the horizontal field $B_{x}/\sqrt{4\pi}$ in $\mu$G. Both panels are for $t=0.37$ Gyr.\\}
\label{BzBx45}
\end{figure} 
\indent
In the face-on case (left panel), there is less gas stripping for most of the simulation time when the wind is magnetized compared to the pure hydro case. One possible interpretation of this trend is that the formation of magnetic draping layer on the face of the galaxy exposed to the incoming wind slows down the gas removal. This trend holds also in the higher resolution runs (we present higher resolution cases only for the face-on cases as these are easier to simulate; the magnetic draping layer is resolved independently of the disk orientation with respect to the wind direction). The magnetic field in the draping layer is dynamically important and the field that enters the disk is already amplified significantly. While the rate of gas removal is slightly larger in the higher resolution runs, for given physics the low and high resolution runs follow the same evolutionary trends, and the stripping rate is always lower in MHD cases than in the hydro cases. Interestingly, the gas pressure stripping rate appears to be  larger in the MHD case than in the hydro case when the disk is tilted.
One interpretation of this effect is that it could be due to the fields sliding over the disk surface exposed to the wind. These sliding fields could ``scrape'' the gas from this surface. Note that this effect would operate across most of the disk surface in the tilted disk case, and that it would not be as efficient in the face-on case where the slipping of the lines occurs only close to the disk edge. We should note that we are using a simplified setup in which the ICM is threaded 
by a uniform magnetic field, and we ignored any internal galactic magnetic fields. Such internal fields may tend to protect the galaxy 
against stripping. Further analysis for a range of wind 
and galaxy parameters, and accompanied by a careful convergence study (e.g., in smaller computational volumes), is needed to test our hypothesis
that the magnetic field may accelerate the gas removal in some cases, or at least not necessarily impede it.\\

\subsection{Formation of the ordered field ahead of the galaxy and filamentary field in the tail}

\begin{figure*} 
\begin{center}
\includegraphics[scale=0.40]{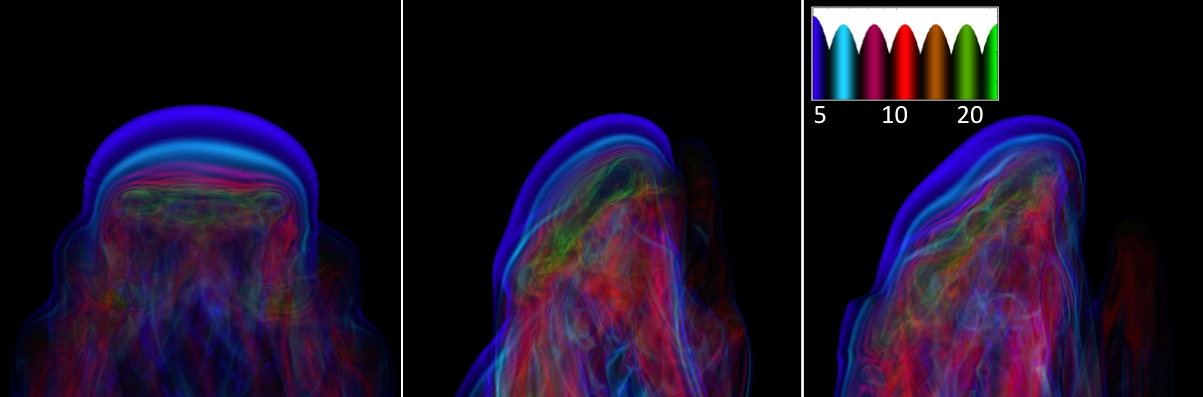} 
\end{center}
\caption{Zoom-in on the magnetic field contours in the vicinity of the galactic disk. Left: disk face-on with respect to the wind. Middle: tilted disk (Case A). Right: tilted disk (Case B). Inset in right panel shows the transfer function used in the volume rendering of the magnetic field (B-field is measured in $\mu$G). Detailed structure of the magnetic field is clearly visible in the electronic version. Each panel is 60 kpc on a side.\\}
\label{Caps}
\end{figure*}

\begin{figure} 
\begin{center}
\includegraphics[scale=0.35]{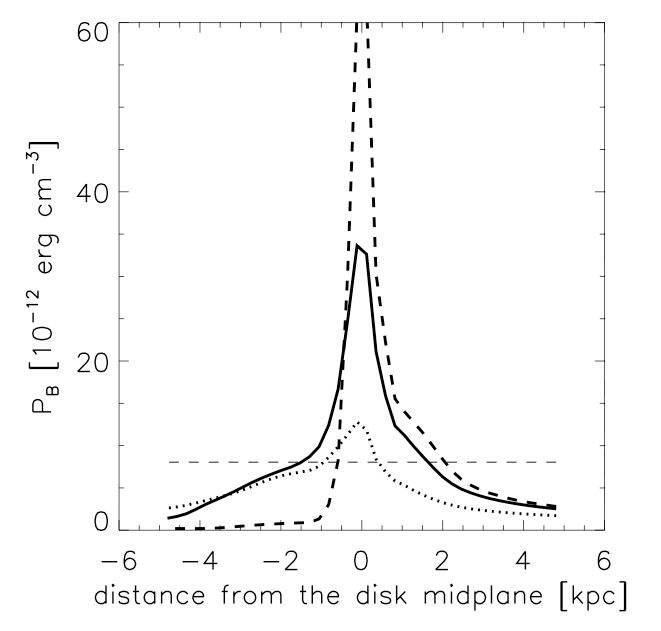} 
\end{center}
\caption{Profiles of the magnetic pressure $P_{B}$ across the galactic disk plane. Front side of the galaxy
corresponds to the positive values of the distance from the disk midplane. $P_{B}$ is computed within cylinders 20 kpc (dotted), 
10 kpc (solid), and 5 kpc (dashed) in radius. Thin horizontal dashed line corresponds to $0.5\rho_{\rm wind}v^{2}_{\rm wind}=0.5P_{\rm ram}$. 
The thickness of the draping layer corresponds to the point where  $P_{B}=0.5P_{\rm ram}$.\\}
\label{Layer}
\end{figure}

In Figure~\ref{BzBx45} we present cross sections ($y=0$) through the computational domain showing the absolute value of $B_{z}$ (left panel) and $B_{x}$ at $t=0.37$ Gyr. The vertical magnetic field reveals the presence of elongated filaments that extend to large distances from the disk of the galaxy.
This figure clearly shows the detached shock in front of the disk side exposed to the wind, which is consistent with the fact that the galaxy moves through the ICM with a slightly supersonic velocity.\\
\indent
A draping layer is established closer to the disk plane than the shock standoff distance. This layer is also clearly seen in Figure~\ref{Caps}. This figure shows
a zoom-in on the magnetic field contours in the vicinity of the galactic disk. In the left panel, the disk is oriented face-on with respect to the wind. Middle and right panels show the tilted disk Case A and Case B, respectively. Inset in right panel shows magnetic field levels in $\mu$G.
The thickness of the draping layer may be defined as the width of the region where the sum of the gas pressure and magnetic pressures
become comparable to the ram pressure.\\
\indent
In Figure~\ref{Layer} we show the profiles of the magnetic pressure $P_{B}$ across the galactic disk plane at $t=0.37$ Gyr for the face-on case. 
The magnetic pressure is computed within cylinders of radii 20 kpc (dotted), 10 kpc (solid), and 5 kpc (dashed). 
Thin horizontal dashed line corresponds to $0.5\rho_{\rm wind}v^{2}_{\rm wind}$. 
The thickness of the draping layer corresponds to the point where  $P_{B}=0.5\rho_{\rm wind}v^{2}_{\rm wind}$. 
This definition is comes from the assumption of the equipartition between the magnetic and thermal pressures in the layer. This assumption is 
better when radiative cooling is slow and we verified that it approximately holds in the draping layer in front of the galactic disk plane.

\begin{figure*} 
\begin{center}
\includegraphics[scale=0.4]{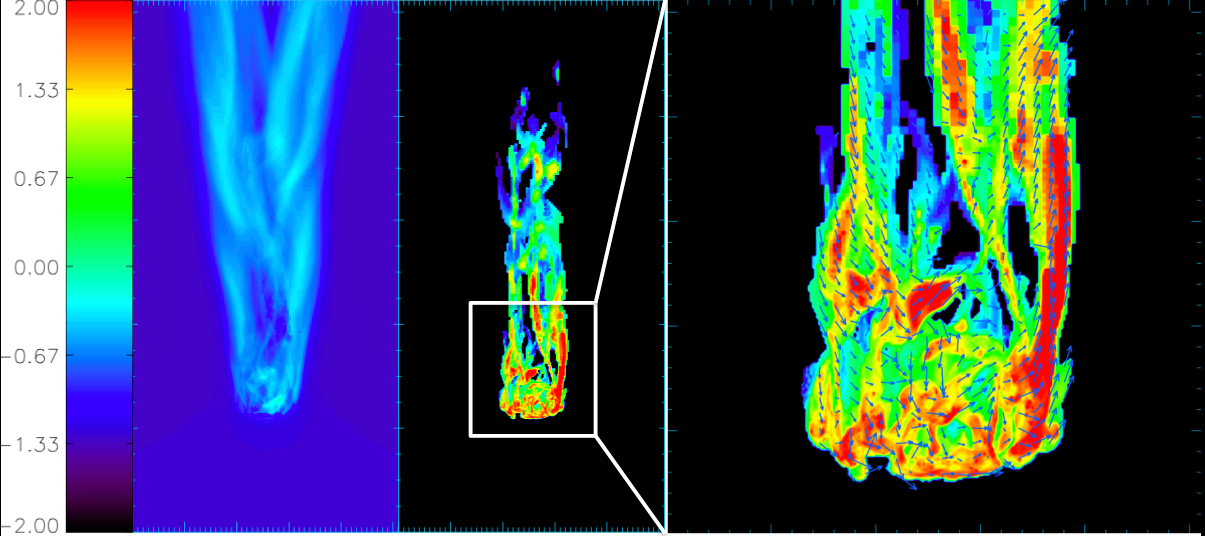} 
\end{center}
\caption{Line-of-sight projections of the ratios of the magnetic and thermal pressures (in log space).
Results for gas temperature more (less) than $10^7$K are shown in left (middle) panel.
Both cases correspond to $t=0.7$ Gyr.
Right: zoom-in on the regions in the vicinity of the disk plane shown in the middle panel.
Magnetic field vectors are superimposed on the image and their length is proportional to the field strength. The size of the left and middle panels is 120 kpc by 240 kpc.\\}
\label{presRatio_zoom}
\end{figure*}

\indent
As can be seen in Figure~\ref{Layer}, the thickness of the layer decreases with the distance from the rotation axis of the galaxy. 
For a spherical obstacle of radius $r$, the theoretical thickness of the draping layer along the line
connecting the observer and the stagnation point is $l\sim R/M_{A}^{2}$ \citep{lyutikov06, dursi08, pfrommer10}. However, the geometry of the problem is different in our case. A finite disk may be better approximated
by a part of a sphere of very large curvature radius $R>r$, and so the thickness of the draping layer is much larger in the vicinity of a disk than close to a spherical obstacle of radius $r$.
For a finite disk, the central thickness of the magnetic layer can be approximated following \citet{erkaev}.
From magnetic freezing, and assuming one-dimensional flow near the flat disk, 
we get $B_{\rm wind}v_{\rm wind}=B(z_{d})v(z_{d})$, where $z_{d}$ is the distance from the galactic plane, and $v$ is the gas velocity normal to the disk plane.
It can be shown that the velocity linearly approaches zero as a function of the distance from any surface along the stagnation line
\citep{landau}. Specifically, 
$v=a\mu v_{\rm wind}$, where $a$ is a dimensionless constant around unity that weakly depends on the surface geometry (e.g., $a=3$ for a sphere, $a=2$ for a cylinder; assumed equal to unity below), 
and $\mu$ is the
dimensionless distance (ratio of the distance to the characteristic size of the obstacle). Thus, the magnetic field is $B(z_{g})=B_{\rm wind}/\mu$.
Comparing magnetic pressure to half the ram pressure, and assuming equipartition and neglecting cooling, we get

\begin{eqnarray}
\frac{B_{\rm wind}^{2}}{8\pi a^{2}\mu^{2}} = \frac{1}{2}\rho_{\rm wind}v^{2}_{\rm wind}\\ \nonumber
\end{eqnarray}

\noindent
The dimensionless thickness of the layer is thus $\mu=M_{A}^{-1}$, which is much larger than
the estimate for a sphere. For a characteristic size $R\sim 10$ kpc of the disk, the central thickness is $l\sim R/M_{A}\sim 2.3$ kpc. 
Close to the edge of the disk the thickness of the layer is expected to be smaller than this rough estimate due to the smaller effective curvature radius
of the surface and the slippage of the magnetic field lines past the disk. These trends are supported by the results shown in Figure~\ref{Layer},
that clearly shows that the typical thickness is around $\sim 2$ kpc (measured within 10 kpc radius cylinder; solid line) which is comparable to 
our estimate. As expected, the layer thickness decreases away from the rotation axis, and is easily resolved in our simulations.
This magnetic layer forms on a timescale $t_{\rm form}\sim l/v_{\rm wind}\sim R/(v_{\rm wind}M_{A})\sim 1.7$ Myr. Note also that, in the tilted disk case, the magnetic layer is thinner on the leading edge of the galaxy exposed to the incoming ICM wind than along the rotation axis in the face-on case. However, even in the tilted disk case, this layer is still resolved in our simulations.\\
\indent
The build-up of the inhomogeneous magnetic field and density inhomogeneities close to the disk (below the draping layer) causes pressures that undo the 
protective effect of the draping layer and allow the gas to stream out of the galaxy. 
In the process, the ordered horizontal fields are transformed into vertical filamentary fields that fluctuate on small scales (development of instabilities is discussed in the next section). Small-scale vertical fields are seen very close to the disk plane in the left panel in Figure~\ref{BzBx45} and in Figure~\ref{Caps}.
In the downstream region, vertical fields are formed that stretch over large distances from the galactic disk.
The amplification of the magnetic field near the disk plane is clearly seen in Figure~\ref{Layer}.
This is caused by the combination of the flux freezing, gas compression due to ram pressure of the 
supersonically moving galaxy, radiative cooling, the 
differential motions in the disk, and the shearing between the disk and the ambient ICM. The increase in the gas density is due to the ram pressure of the wind and 
the radiative cooling which allows for the gas condensation in the disk plane.

\begin{figure} 
\begin{center}
\includegraphics[scale=0.30]{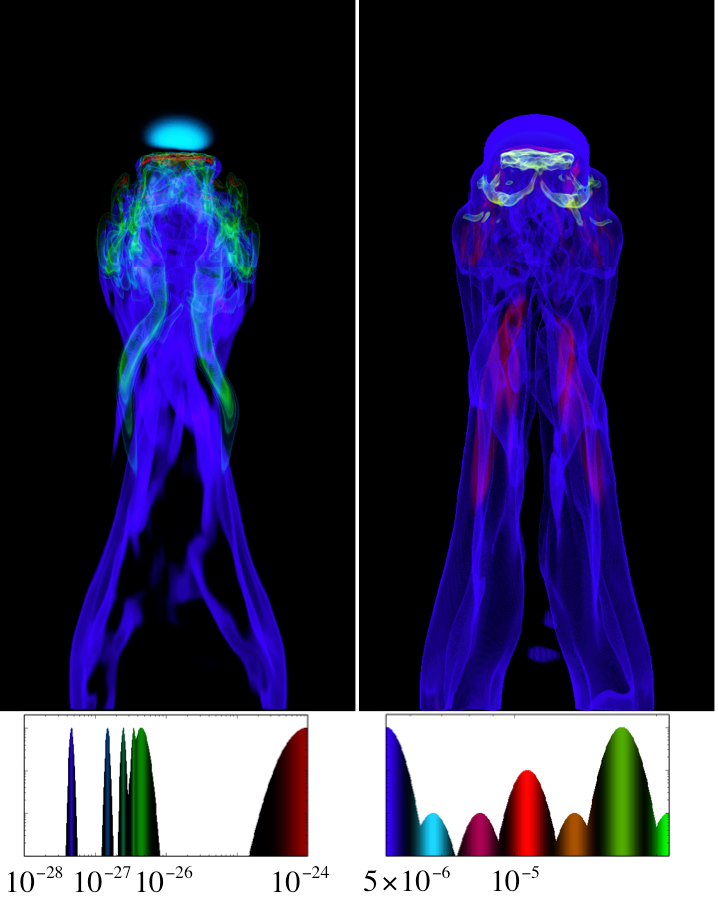} 
\end{center}
\caption{Volume rendering of the gas density (left) and magnetic field strength for the disk face-on with respect to the ICM wind. 
The disk of the galaxy is seen edge-on. The snapshot is taken at 0.37 Gyr. 
The iso-surfaces are 
displayed using the volume rendering transfer function shown
below the images. The unit of density is g cm$^{-3}$ and the magnetic field is in Gauss. Double tail in the magnetic field is clearly seen in the right panel. Double density tails, with densities larger than the wind density, extend up to the middle of left panel (shown in light green;  see electronic version of the figure).\\}
\label{tails}
\end{figure}
 
\begin{figure*} 
\begin{center}
\includegraphics[scale=0.40]{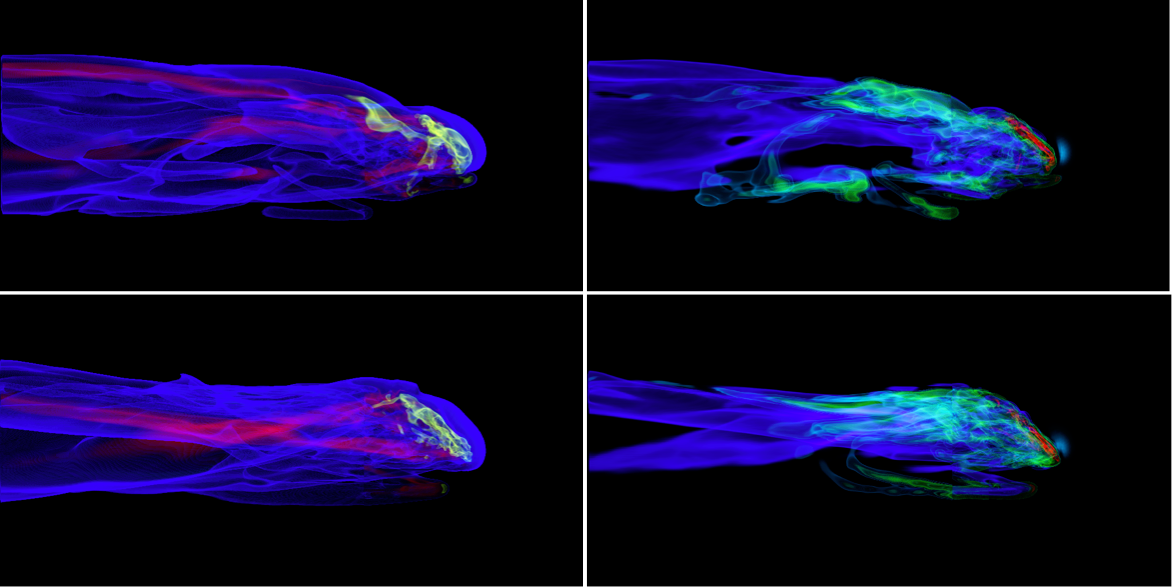} 
\end{center}
\caption{Volume rendering of the magnetic field strength (left column) and the gas density. The disk of the galaxy is seen edge-on and the snapshots are taken at 0.37 Gyr. Top row is for Case A and bottom one for Case B. The iso-surfaces and units are the same as in Figure~\ref{tails}. The size of each panel is 240 kpc by 120 kpc.\\}
\label{f9}
\end{figure*} 

\begin{figure*} 
\begin{center}
\includegraphics[scale=0.40]{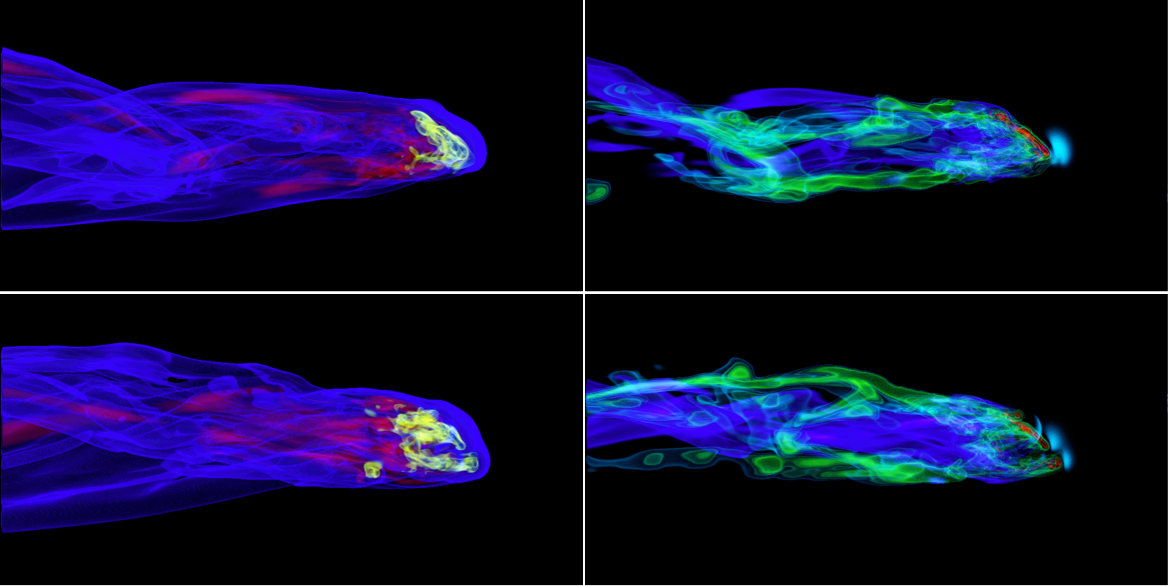} 
\end{center}
\caption{Same as Figure~\ref{f9} but for $t=0.47$ Gyr.\\}
\label{f10}
\end{figure*}

\subsection{Disk instabilities}
The velocity gradient between the disk and the ICM, tangles the field close to the disk midplane and below the draping layer.
Tangling is seen in both the $B_{z}$ and $B_{x}$ maps (left and right panels in Figure~\ref{BzBx45}), and in Figure~\ref{Caps}, and it is caused in part by shear instabilities whose growth rate is given by 

\begin{eqnarray}
\omega_{\rm KH} = \frac{2\pi}{\lambda} \Delta v\frac{\sqrt{\rho_{\rm ism}\rho_{\rm icm}}}{\rho_{\rm ism}+\rho_{\rm icm}}, \\ \nonumber 
\end{eqnarray}

\noindent
where $\lambda$ is the wavelength of the unstable mode and $\Delta v$ the velocity difference across the shear layer. Taking $\rho_{\rm ism} \gg \rho_{\rm icm}$, we can write for the e-folding growth time

\begin{eqnarray}
t_{\rm shear} =\omega_{\rm KH}^{-1}\approx 40 {\rm Myr}\, \frac{\lambda}{1\, {\rm kpc}}  \frac{\rm 200\, km/s}{\Delta v} \frac{\sqrt{\rho_{\rm ism}/\rho_{\rm icm}}}{50}. \\ \nonumber
\end{eqnarray}

\noindent
We note that the shearing layer may in principle be stabilized to some extent by the disk gravity. 
However, the Richardson number $Ri$ that measures the ratio of potential to kinetic energy is

\begin{eqnarray}
Ri\sim\frac{g_{z}h}{\Delta v^{2}}\sim\left(\frac{h}{r}\right)^2, 
\end{eqnarray}

\noindent
where $h$ is the disk thickness, $g_{z}$ is the vertical component of the gravitational field in the disk, and where we assumed that 
the shearing velocity $\Delta v$ is comparable to the orbital velocity. For tilted disk case, the relevant shear velocity will be even larger because the wind speed is larger than the orbital speed. Since the disk is thin, $Ri\ll 1$ and gravity will not prevent the 
KH instability. However, the KHI can also be suppressed by the finite thickness of the shearing layer. General arguments suggest 
that modes $kd\ga 1$, where $d$ is the thickness of the shearing layer, are suppressed \citep{chandra,ferrari,chi}. For $d\sim 1$ kpc, 
perturbations characterized by 
$\lambda\la$ a few kpc should be suppressed. This sets a lower limit on $\lambda$ in Eq. (9). 
We also point out that the growth of the KH unstable modes will be limited by the onset of nonlinear saturation, and the modes much shorter
than the disk thickness are unlikely to significantly perturb the disk.
Magnetic fields will suppress unstable modes parallel to the field but will not suppress modes perpendicular to the field lines
\citep{dursi07}. Thus, in general, the shearing instability due to the shearing motions between the galaxy and the ICM will not be suppressed by the magnetic fields.\\
\indent
The galactic disk can also be unstable to the Rayleigh-Taylor instability (RTI). The infalling galaxy is subject to the ICM drag force

\begin{eqnarray}
F_{\rm drag}=\frac{1}{2}c_{p}\rho_{\rm wind}v_{\rm wind}^{2}A,
\end{eqnarray}

\noindent
where $c_{p}$ is the cofficient of parasite drag, $\rho_{\rm wind}=\rho_{\rm icm}$ 
and $A$ is the cross-section of the obstacle \citep{vonMises,batchelor}. For a circular disk oriented face-on with respect to the wind 
$c_{p}=1.11$. When magnetic layer is formed the magnetic tension dominates over turbulent drag and the effective drag force is larger by a factor of order unity. For example, for a sphere, \citet{dursi08} find $c_{p}=1.87$.
This drag force could cause the RTI. However, as pointed out by several authors (e.g., \citet{elke} or \citet{murray93}), 
the gravitational acceleration $g_{\rm disk}$ of the disk can suppress the RTI if $g_{\rm disk}Ah\rho\ga F_{\rm drag}$, which leads to 
the same condition as that for the absence of the ram pressure stripping, namely $g_{\rm disk}\Sigma_{\rm disk}\ga P_{\rm ram}$. Thus, 
the gas does not become RT unstable before it gets stripped. 

Even the stripped gas may potentially be stabilized against RTI by the self-gravity of the gas if

\begin{eqnarray}
2\pi G\Sigma_{\rm disk}\ga\frac{c_{p}\rho_{\rm icm}v^{2}_{\rm wind}}{2\rho_{\rm ism}h}.
\end{eqnarray}

\noindent
This implies that the stripped gas is stable against RTI if the surface density exceeds $\Sigma_{\rm crit}$ given by

\begin{eqnarray}
\Sigma_{\rm crit}\sim \left(\frac{c_{p}}{4\pi}\frac{P_{\rm ram}}{G}\right)^{1/2}.
\end{eqnarray}

\noindent
and the value of $\Sigma_{\rm crit}$ is smaller than $\Sigma_{\rm disk}$ in our simulations. 

Finally, magnetic fields may also stabilize the growth of the perturbations along the field lines
if the perturbations correspond to wavelengths smaller than a critical length \citep{chandra,dursi08}. 
However, perturbations in the direction perpendicular to the magnetic field will not be suppressed. Three-dimensional simulations demonstrate that 
the overall RTI growth rate will not be supressed and may in fact be somewhat accelerated in the MHD case \citep{stonegardiner}. \\
\indent
The timescale of the gravitational instability, $t_{\rm grav}$, 
is comparable to the orbital timescale $t_{\rm orb}=r/v_{\rm gal}\sim 100\; {\rm Myr}(r/20{\rm kpc})(v_{\rm gal}/100{\rm km/s})^{-1}$.
We also point out that the formation timescale $t_{\rm form}$ of the draping layer is shorter than all of the above timescales, so the hierarchy of 
timescales is: $t_{\rm form}<t_{\rm shear}\la t_{\rm grav}$. All of these timescales are significantly shorter than the simulation time.
The estimates presented above are consistent with the observation that no clear disk fragmentation occurs before $\sim 50$ to 100 Myr 
when the wind reaches the disk, and with the fact that the disk is clearly unstable at $\sim 0.37$ Gyr.

\subsection{Magnetic pressure support and field orientation in the tail filaments}

Figure~\ref{presRatio_zoom} shows the line-of-sight projections of the ratios of the magnetic and thermal pressures at 0.7 Gyr.
Results for the gas temperature more (less) than $10^7$K are shown in left (middle) panel.
The right panel shows a zoom-in on the regions in the vicinity of the disk plane shown in the middle panel.
Magnetic field vectors are superimposed on the image and their length is proportional to the field strength.
Magnetic fields tend to be aligned with those regions where the magnetic pressure support dominates over thermal pressure.
This hints at the interesting possibility that this effect
could prevent anisotropic thermal conduction from evaporating the cool magnetically supported filaments.

\subsection{Formation of double tails}
In Figure ~\ref{tails} we show the volume rendering of the gas density (left panel) and magnetic field strength. The disk of the galaxy is seen edge-on
in both panels and the snapshot is taken at 0.37 Gyr. The iso-surfaces used in the rendering are shown immediately below the images. 
The right panel in this figure clearly illustrates that the magnetic field is ordered on the working surface of the galactic disk 
(i.e., in front of the galactic disk) and that 
the field gets tangled immediately behind the disk and then is stretched along the direction of the wind. Two magnetic tails are clearly seen behind the galaxy in the
right panel. While the gas distribution is in general filamentary, it is clear that 
two dominant gaseous tails are present and extend up to the middle of the image in the z-direction (shown in light green; most easily visible in the electronic version of this figure). These tails are spatially correlated with the magnetic tails and are denser than the ambient material that has density comparable to that of the wind density.
We emphasize that the tails are not due to ``limb brightening'' of a hollow cone of the gas density or magnetic field distribution seen in projection onto the 
plane of the sky. Face-on projections (not shown) demonstrate that the two separate tails lie along the diagonal $x-y$ plane, which is consistent with the direction of the magnetic field in the incoming ICM wind.
However, at later times the two distinct tails get washed out or they are not as cleanly separated. \\
\indent
The impact of the galaxy interaction time with the ICM and the disk inclination with respect to the incoming ICM wind is shown in Figures~\ref{f9} and ~\ref{f10}.
In both figures the disk of the galaxy is seen edge-on. The snapshots in Figures~\ref{f9} and \ref{f10} are taken 0.37 Gyr and 0.47 Gyr, respectively. 
Given the wind speed of 1300 km/s, the time difference of 0.1 Gyr between the results shown in Figures~\ref{f9} and ~\ref{f10} translates to 55\% of the longer box length $L_{z}$ ($L_{z}=240$ kpc). Top rows correspond to Case A and bottom ones to Case B. Left columns present magnetic field rendering and right ones the gas density. The iso-surfaces and units are the same as in Figure~\ref{tails}. This figure confirms that the magnetic field has a strong effect on the morphology of the ram pressure stripping tails. The change from Case A to Case B in the magnetic field orientation in the incoming ICM wind leads to marked morphological differences in the tail magnetic field and gas density. In all cases, the gas density and the distribution of the magnetic field appears fluctuating but generally quite filamentary. \\
\begin{figure*} 
\begin{center}
\includegraphics[scale=0.40]{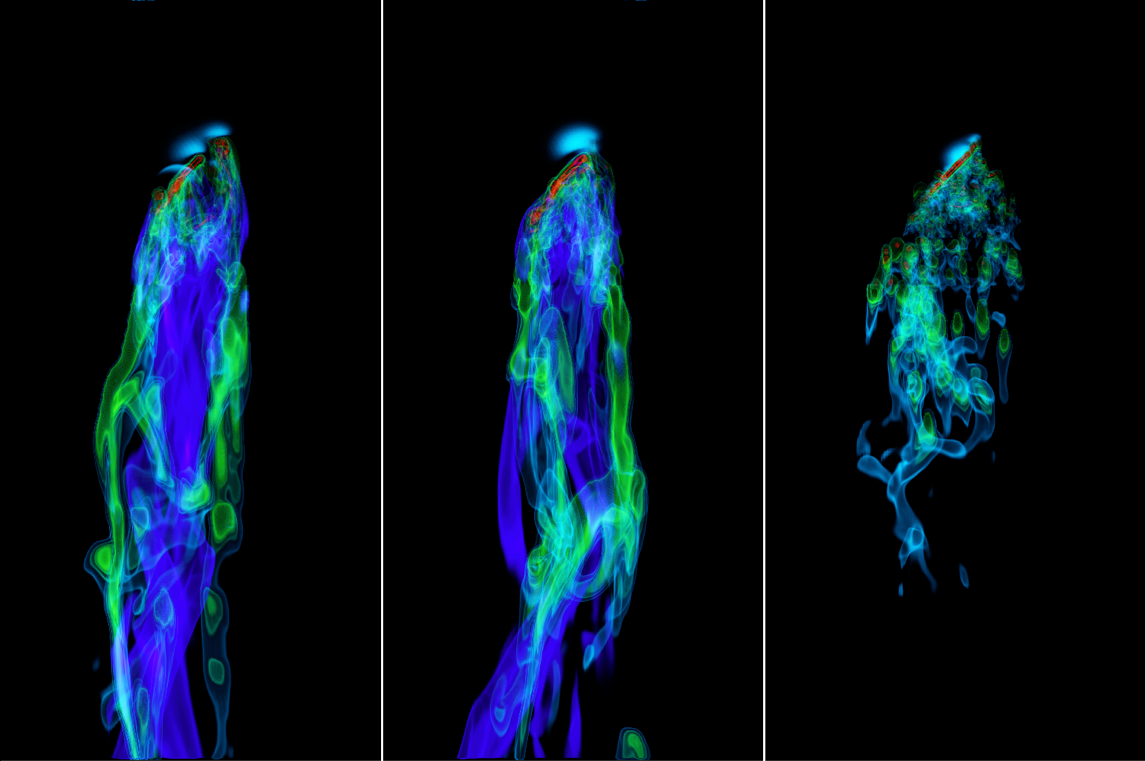} 
\end{center}
\caption{Volume rendering of the gas density at $t=0.47$ Gyr. 
The columns correspond to Case B, Case A, and pure hydro case, from left to right. Density contours 
are at the same levels as in Figure~\ref{tails}. The size of each panel is 120 kpc by 240 kpc.\\}
\label{f11}
\end{figure*} 
\indent
Figure~\ref{f11} combines volume rendering of the gas density for two different orientations of the incoming field (left: Case B; right: Case A) and for the pure hydro case (rightmost panel). All panels correspond to $t=0.47$ Gyr and the galactic disk tilted by 45 degrees with respect to the direction of the galactic motion. The difference between the MHD cases and the pure hydro case is striking.  While the magnetized wind case leads to filamentary tails, the tail in the hydro case is very clumpy. Even though the gas density in the tails fluctuates in time and space in the MHD case, the gas tends to form distinctive filamentary tails that may appear double depending on the orientation of the line-of-sight.  The gas tends to be removed from the disk along the directions approximately within the plane containing the galactic spin vector and the direction of the galactic motion through the ICM. That is, one tail is anchored at leading edge of the disk exposed to the ICM wind and the other one originates from the gas sliding along the disk surface and is anchored at the trailing edge of the disk. This situation is different from that presented in Figure~\ref{tails}, that shows a face-on disk, where the tail orientation is determined predominantly by the direction of the incoming ICM field. In the tilted disk case, the morphology of the tails is determined by the combination of the disk inclination and the orientation of the magnetic fields in the wind rather than just by the orientation of the magnetic field in the wind. Which mechanism for shaping the tails dominates depends on a complex
interplay between the time since the galaxy enters the ICM, magnetic field orientation, and the disk tilt. For example, for disk inclinations smaller than 45 degrees
the impact of the ICM field orientation should be progressively more important in shaping the tail properties than the combined effect of the disk tilt and magnetic forces.
The same is generally true also for earlier times since the galaxy enteres the cluster atmosphere. The recently detected double tails in ESO 137-002 have been suggested by \citet{zhang13} to be associated with the early stages of the infall of a disk galaxy into the cluster atmosphere (though the authors also suggested that the disk tilt is large).\\
\indent
We considered incoming ICM fields at inclinations of +/- 45 degrees (Cases A and B, respectively) with respect to the x-axis in the x-y plane. Since the magnetic field has a substantial effect on the tail morphology, the 3D structure of the tails and the overall morphology of the sky-projected gas stripped from the galaxy may be different for ICM fields oriented entirely along the x-axis, y-axis, when the ICM has a non-vanishing z-component, or when internal fields are included. For example, recall that for the face-on disk case, the dominant tails shown in Figure~\ref{tails} are intrinsically narrow and located in the plane of the incoming field (despite the fact that the fields are sliding in the x-y plane and along the direction perpendicular to the direction of the incoming field). This suggests that, even in the tilted disk case, and when the incoming field has a direction closer to the intermediate case between Case A and B (i.e., magnetic field closer to the x-direction; galaxy tilted about the y-direction), the intrinsic width of each tail may be narrower. Full study of all these possibilities requires additional (and very CPU-intensive) runs and is beyond the scope of this initial investigation of the impact of the magnetic fields on the ram pressure stripping process.
Here we merely point out that the dynamical interaction of the wind and the galaxy may naturally lead to formation of 
tails that appear as bifurcated in the plane of the sky and that such tails can be formed under a 
variety of situations, both for the disks oriented face-on with respect to the ICM wind or for tilted ones. This is possible due to the generic tendency of the magnetic fields to produce intrinsically filamentary (rather than clumpy) structures that may be characterized by two dominant filaments. We emphasize that no such effects are seen in the pure hydro cases.
This proof-of-concept result suggests that the stripping process could generate observable double tails. 
Such a double tail behind a galaxy has indeed been observed in ESO 137-001 (Sun et al. 2006, 2007) and in ESO 137-002 \citep{zhang13}.
It is intriguing that double tails have been observed in both nearly face-on (Sun et al. 2007) and in significantly tilted disks \citep{zhang13}, and our results are broadly consistent with these observations.
We also note that the density contrast between the tails and the ambient ICM depends on a number of factors including the density of the ICM wind, magnitude of the ram pressure, time since the galaxy entered the cluster, and the depth of the galactic potential well. This contrast could be higher (or lower) depending on these factors. In an followup study we will present observational diagnostics -- HI, H$\alpha$, X-ray signatures -- for a  
range of the parameters characterizing the wind and galaxy, including disk orientation and ICM and intrinsic galactic fields.

\subsection{Phase space distribution for the gas and the magnetic field}
\begin{figure*} 
\begin{center}
\includegraphics[scale=0.40]{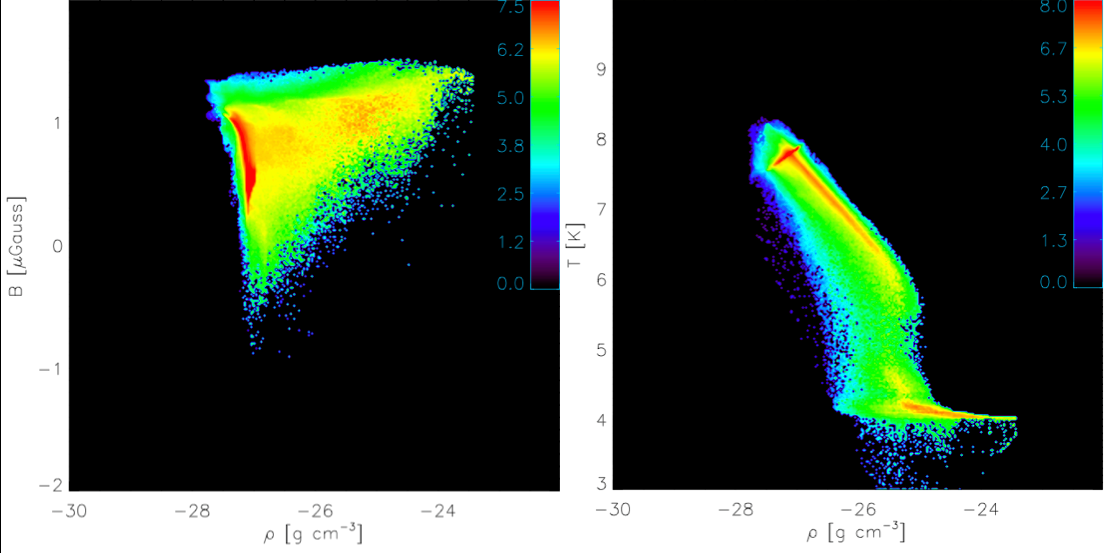}
\end{center}
\caption{Left: $B-\rho$ phase space for the gas downstream from the galaxy. 
Right: $T-\rho$ phase space for the gas downstream from the galaxy (MHD case). Both panels correspond to $t=0.7$ Gyr, face-on orientation fo the disk, and the MHD case. Both the magnetic field and temperature are in log space. Color bars show the gas mass in solar masses equally spaced in log space between 0 and 7.5 (left panel) and 0 and 8.}
\label{f12}
\end{figure*}

The left panel in Figure ~\ref{f12} shows the $B-\rho$ phase space for the gas downstream from the galaxy and the right one presents the corresponding $T-\rho$ phase space for the gas. Both panels correspond to $t=0.7$ Gyr, face-on orientation of the disk, and the MHD case.
The color bar shows the amount of gas in solar masses in each $B-\rho$ or $T-\rho$ bin. 
The left phase space diagram demonstrates that there are no high density$-$low magnetic field strength regions. The main peak in the $B-\rho$
corresponds to the relatively hot ICM. The magnetic fields in this gas can be comparable to or higher than those in the incoming ICM wind.
The secondary high magnetic/density peak in the left $B-\rho$ distribution corresponds to the galactic gas. This secondary peak in the $B-\rho$ distribution
corresponds to systematically higher magnetic fields than those belonging to the left peak. 
The right panel reveals that the gas is clearly multiphase. 
Significant amounts of gas near $T\sim 10^{7}$K and $10^{-27}$ g cm$^{-3}$ are due to the wind and some evidence for adiabatic gas compression is clearly seen near this location in the phase space.
Comparison of the left and right panels shows that this secondary peak in the $B-\rho$ distribution corresponds to relatively cool gas. As discussed above, the distribution of this gas phase is very filamentary. 
Both the stretching of the nearly incompressible gas, as well as the radiative cooling that increases the gas density, bring the denser gas and magnetic fields into close contact. While numerical mixing-in of the magnetic field into the stripped gas makes it appear that the densest and coolest gas in the tail is always relatively strongly magnetized,
in an infinite resolution ideal MHD case, the strong magnetic fields and the dense gas may not be fully mixed. However, they should nevertheless be confined to the same non-volume filling locations in space. That is, these regions/filaments, while not having uniform distribution of density and magnetic field inside them, should on average be more dense and magnetized than their surroundings. \\

\section{Conclusions}

We performed hydrodynamical and magnetohydrodynamical simulations of ram pressure stripping in a disk galaxy including radiative cooling and self-gravity of the gas. Our main conclusions can be summarized as follows:

\begin{enumerate}

\item The presence of ambient magnetic fields has a strong impact on the morphology of the tail. While the MHD case shows long ($> 100$ kpc) filamentary structures in the tail, in the purely hydrodynamical case the tail is very clumpy. 

\item In the filaments of the ram-pressure stripped tail the magnetic pressure support dominates and the field tends to be aligned with the filaments. This may suppress thermal conduction between filaments and the hot ambient ICM and could be relevant to the question of whether stars can form in the ram pressure stripped gas. The filamentary structure helps to explain observations of ram-pressure stripped tails. Our finding that the galactic wake consists of long filaments that are magnetically supported helps to explain observations such as those by Yoshida et al. (2008) who find narrow, blue filaments with H$\alpha$ emission that extend up to 80 kpc from a galaxy in the Coma cluster. 

\item The ram pressure stripping process may lead to the formation of
tails that appear as bifurcated in the plane of the sky. Such tails can be formed under a 
variety of situations, both for the disks oriented face-on with respect to the ICM wind, and for tilted ones. 
This bifurcation is the consequence of the generic tendency for the magnetic fields to produce very filamentary tail morphology. 
The tail morphology depends on a number of factors including the tilt of the galaxy with respect to the galaxy's direction of motion and the orientation of the magnetic field around the galaxy.
Interestingly, double tails have recently been observed in near face-on and significantly tilted disks \citep{sun06,sun07,zhang13}.
The detectability of such filamentary structures depends on a number of factors: the 
time since the beginning of the stripping process, the characteristic lengthscale of the magnetic field fluctuations in the ICM or the inclination of the ICM field with respect to the galaxy,
the orientation of the galaxy with respect to the line-of-sight, disk tilt with respect to the galaxy's direction of motion, depth of the galactic potential well, and the relative emissivities of the tail and the ambient ICM.

\item The impact of the magnetic field on the magnitude of the ram pressure gas removal from the galaxy is generally weak.
The presence of uniform magnetic field in the ICM wind interacting with a face-on galaxy tends to reduce the amounts of gas stripping compared to the pure hydrodynamical case. This may be due to the formation of a stable magnetic draping layer on the side of the galaxy exposed to the incoming ICM wind.
For significantly tilted disks the situation may be reversed and the stripping rate may be larger when magnetic fields are present in the ICM. This may be caused by the ``scraping'' of the disk surface by the magnetic fields sliding past the ISM/ICM interface. The impact of internal galactic fields also needs to be investigated in future work. Moreover, the impact of the wind magnetic fields tangled predominantly on scales significantly smaller than the disk size also remains to be explored.

\item As the gravitational instability in the disk and the disk shearing against the ambient ICM are not suppressed by the magnetic fields.
Consequently, the ordered horizontal fields are transformed into vertical filamentary fields that fluctuate on small scales. 
The impact of magnetic fields on the ram pressure stripping may be different in giant elliptical galaxies where the rotation does not play a significant role and where the absence of dense disk changes the flow geometry of the stripped gas.

\end{enumerate}

\section*{Acknowledgments}
The software used in this work was in part developed by the DOE NNSA-ASC OASCR Flash Center at the University of Chicago.
MR acknowledges the NSF grant NSF 1008454 and NASA ATP 12-ATP12-0017.
MB acknowledges support by the research group FOR 1254 funded by the Deutsche Forschungsgemeinschaft. 
We thank the Kavli Institute for Theoretical Physics at University of California, Santa Barbara where part of this work was performed.
We thank Fabrizio Brighenti for the routine to compute the flattened King potential
and Ming Sun, Andrey Kravtsov, Joel Bregman, Eric Bell, Eugene Churazov, Justin Nieusma, Christoph Federrath and John ZuHone for insightful discussions. 
We thank Mitch Begelman for help with securing access to computational resources in the initial stages of this project.
Simulations were performed on the {\it Pleiades} machine at NASA
Ames (grant number SMD-11-2346; PI Ruszkowski) and on the {\it Galaxy} cluster at the University of Michigan. 
We thank Tim Sandstrom at NASA for his help with volume rendering of the simulation data. MR thanks Matt Turk and Sam Skillman for help with the $yt$ visualization software that was used to generate volume rendering images presented in this paper.

\bibliography{stripping} 

\label{lastpage}

\end{document}